\documentclass{jkas}

\def\year{2016} 
\def\month{June} 
\def\volume{49} 
\def\issue{3} 
\def\beginpage{001} 
\def\endpage{010} 
\def\received{March 20, 25} 
\def\accepted{April **, 2016} 
\date{Received \received; accepted \accepted}

\def\kms{~{\rm km~s^{-1}}}
\def\cm3{~{\rm cm^{-3}}}
\def\yr{~{\rm yr}}

\def\muG{~{\mu\rm G}}

\usepackage{flushend}

\title{Re-acceleration model for the ``Toothbrush'' Radio Relic}

\author{Hyesung Kang}

\affil{Department of Earth Sciences, Pusan National University, Pusan 609-735, Korea; \email{hskang@pusan.ac.kr}}

\def\corrauthor{H. Kang}
\def\runningauthor{Kang}
\def\runningtitle{Radio Emission from Weak Spherical Shocks in the Outskirts of Galaxy Clusters}
\def\keywords{acceleration of particles --- cosmic rays ---
galaxies: clusters: general --- shock waves}

\def\abstracttext{
The Toothbrush radio relic associated the merging cluster 1RXS J060303.3 is presumed to be produced 
by relativistic electrons accelerated at merger-driven shocks. 
Since the shock Mach number inferred from the observed radio spectral index, $M_{\rm radio}\approx 2.8$, is larger than
that estimated from X-ray observations, $M_{\rm X-ray}\lesssim 1.5$,
we consider the re-acceleration model in which a weak shock of $M_s\approx 1.2-1.5$ sweeps through the
intracluster plasma with a preshock population of relativistic electrons.
We find the models with a power-law momentum spectrum with the slope, $s\approx 4.6$, and the cutoff Lorentz
factor, $\gamma_{e,c}\approx 7-8\times 10^4$ can reproduce reasonably well the observed profiles of radio fluxes 
and integrated radio spectrum of the head portion of the Toothbrush relic.
This study confirms the strong connection between the ubiquitous presence of fossil relativistic plasma originated from AGNs  
and the shock-acceleration model of radio relics in the intracluster medium.
}

\begin{document}
\jkashead 

\section{Introduction}

Some galaxy clusters contain diffuse radio sources on the scales as large as $\sim 2$~Mpc, called `radio relics' 
\citep[e.g.,][]{feretti12,brug12, brunetti2014}. They could be classified into two main groups 
according to their origins and properties \citep{kempner04}.
{\it AGN relics/Radio phoenix} are thought to originate from AGNs.
AGN relics are radio-emitting relativistic plasma left over from jets and lobes of extinct AGNs,
and eventually turn into `radio ghosts' when the electron plasma becomes radio-quite due to synchrotron and
inverse Compton (iC) energy losses 
\citep{ensslin99}.
Radio ghosts can be reborn as {\it radio phoenix}, 
if cooled electrons with the Lorentz factor $\gamma_e \lesssim 100$ are compressed 
and re-energized to $\gamma_e\sim 10^4$ by the structure formations shocks \citep{ensslin01}.
AGN relics/Radio phoenix typically have roundish shapes and steep-curved integrated spectra 
of aged electron populations, and they are found near their source AGN 
\citep[e.g.,][]{slee01,vanweeren11,clarke13,deGasperin15}.
{\it Radio gischt relics}, on the other hand, are thought to be produced via Fermi first order  
process at merger-driven shock waves in the intracluster medium (ICM).
They show thin elongated morphologies, spectral steepening downstream of the putative shock,
integrated radio spectra with a power-law form, and high polarization level \citep{ensslin98, vanweeren10}.

Most of the observed features of radio {\it gischt} relics can be adequately explained by diffusive shock acceleration (DSA) model
in which relativistic electrons are (re-)accelerated at shock waves:
(1) gradual spectral steepening across the relic width, 
(2) power-law-like integrated spectrum with the spectral index that is larger by 0.5 than the spectral index at the relic edge,
i.e., $\alpha_{\rm int}\sim \alpha_{\rm shock} + 0.5$,
and (3) high polarization level up to 50 \% that is expected from the shock compression of turbulent magnetic fields 
\citep[e.g.,][]{vanweeren10,brunetti2014}. 
Moreover, it is now well accepted from both observational and theoretical studies 
that the cosmological shocks are ubiquitous in the ICM
and nonthermal particles can be accelerated at such shocks via the DSA process
\citep[e.g.,][]{ryu03,vazza09,skill11,hong14}.

Yet there remain some puzzles in the DSA origin of radio gischt relics such as low acceleration efficiency of weak shocks and
low frequency of merging clusters with detected radio relics.
The structure formation shocks with high kinetic energy fluxes, 
especially those induced by cluster mergers, are thought to have small Mach numbers of $M_s\lesssim 3$
\citep{ryu03, vazza09}, while the DSA efficiency at such weak shocks is expected to be extremely low \citep{kangryu11}.
In particular, it is not well understood how seed electrons are injected
into the Fermi first order process at weak shocks that form in the high-beta ($\beta= P_{\rm gas}/P_{\rm mag} \sim 100$) ICM 
plasma \citep[e.g.,][]{kang14}.
According to structure formation simulations, the mean separation between shock surfaces in the ICM is about 1~Mpc \citep{ryu03}.
Then several shocks and associated radio relics could be present in merging clusters, if every shock were to produce
relativistic electrons.
Yet only $\sim 10$ \% of luminous merging clusters are observed to host radio relics \citep{feretti12}.
Furthermore, in some clusters a shock is detected in X-ray observations without associated diffuse radio sources
\citep[e.g.,][]{russell11}.

In order to solve these puzzles
re-acceleration of fossil electrons pre-exiting in the ICM was proposed by several
authors \citep[e.g.,][]{kangryu11, kang12,pinzke13}.
The presence of radio galaxies, AGN relics and radio phoenix implies that the ICM  may host radio-quiet ghosts of fossil electrons
\citep{kangryu16}.
In addition, fossil electrons can be produced by previous episodes of shocks and turbulence that are induced by merger-driven activities.
Since fossil electrons with $\gamma_e\gtrsim 100$ can provide seed electrons to the DSA process and enhance the acceleration efficiency,
they will alleviate the low acceleration efficiency problem of weak ICM shocks.
Moreover, the re-acceleration scenario may explain the low occurrence of radio relics
among merging clusters, since the ICM shocks can light up as radio relics only when they 
encounter clouds of fossil electrons \citep{kangryu16}.

The giant radio relic found in the merging cluster 1RXS J060303.3, the so-called ``Toothbrush" relics
is a typical example of radio gischt relics.
It has a peculiar linear morphology with multiple components that look like the head and handle 
of a toothbrush \citep{vanweeren12}.
Recently, \citet{vanweeren16} reported that the spectral index at the northern edge of B1 relic
(the head portion of the Toothbrush) is
$\alpha_{\rm sh}\approx 0.8$ with the corresponding radio Mach number, $M_{\rm radio}\approx 2.8$ 
(see Equation \ref{alpha}). 
But the gas density jump around B1 inferred from X-ray observations implies a much weaker shock with $M_{\rm X-ray}\approx 1.2-1.5$.
This discrepancy between  $M_{\rm radio}$ and $M_{\rm X-ray}$ can be explained,
if a pre-existing population of relativistic electrons has the right power-law slope, i.e., $f_{up}(p)\propto p^{-4.6}$.

In this study, we attempt to explain the observed properties of relic B1 
by the re-acceleration model in which a low Mach number shock ($M_s\approx 1.2-1.5$) 
sweeps though a cloud of pre-existing relativistic electrons.
In the next section, we explain some basic physics of the DSA model, 
while the numerical simulations and the shock models are described in Section 3.
The comparison of our results with observations is discussed in Section 4.
A brief summary is given in Section 5.

\section{DSA Model}

\subsection{Electron and Radiation Spectra of Radio Relics}

According to the DSA model for a {\it steady planar} shock, the electrons that are injected and accelerated at 
a shock of the sonic Mach number $M_s$ form a power-law momentum distribution function, 
$f_e(p,r_s)\propto p^{-q}$ with the slope $q=4M_s^2/(M_s^2-1)$ \citep{dru83}.
Then the electron spectrum integrated over the downstream region of the shock steepens by one power of momentum, i.e.,
$F_e(p)\propto p^{-(q+1)}$ due to synchrotron and iC cooling \citep{ensslin98}. 

The radio synchrotron spectrum radiated by those electrons becomes a power-law of
$j_{\nu}(r_s)\propto \nu^{-\alpha_{\rm sh}}$ at the shock position with the `shock' index 
\begin{equation}
\alpha_{\rm sh}={{(q-3)}\over 2 }= {{(M_s^2+3)}\over {2(M_s^2-1)}}.
\label{alpha}
\end{equation}
This relation is often used to infer the Mach number of the putative shock, $M_{\rm radio}$,
from the radio spectral index.
Moreover, the volume-integrated radio spectrum behind the shock becomes another steepened power-law, $J_{\nu} \propto \nu^{-A_{\nu}}$ 
with the `integrated' index $A_\nu=\alpha_{\rm sh}+0.5$ above the break frequency $\nu_{\rm br}$.
Note that the break frequency depends on the magnetic field strength, $B$, and the shock age, $t_{\rm age}$, as follows:
\begin{equation}
\nu_{\rm br}\approx 0.63 {\rm GHz} \left( {t_{\rm age} \over {100 \rm Myr}} \right)^{-2}
 \left[ {(5\muG)^2} \over {B^2+B_{\rm rad}^2}  \right]^{2} \left( {B \over {5\muG}} \right),
\label{fbr}
\end{equation}
where $B_{\rm rad}=3.24\muG(1+z)^2$ and $B$ is expressed in units of $\muG$
\citep{kang11}.

However, such expectation for simple power-law spectra needs to be modified in real observations.
If the break frequency $\nu_{\rm br}$ lies in the range of observation frequencies (typically $0.1\lesssim \nu_{\rm obs} \lesssim 10$~GHz),
for instance, the integrated radio spectrum steepens gradually over $\sim(0.1-10)\nu_{\rm br}$,
instead of forming a single power-law \citep{kang15a}.
Furthermore, in the case of spherically expanding shocks with varying speeds, 
both the electron spectrum and the ensuing radio spectrum exhibit spectral curvatures \citep{kang15b}.
In fact, the spectral steepening above $\sim 2$~GHz have been detected in the relic in A2256 \citep{trasatti15}
and the Sausage relic in CIZA J2242.8+5301 \citep{stroe16}.
The integrated spectral index of B1 of the Toothbrush relic increases from $A_\nu \approx 1.0$ 
below $2.5$~GHz to $A_\nu \approx 1.4$ above $2.0$~GHz \citep{stroe16}.

In the re-acceleration model, on the other hand,
the re-accelerated electron spectrum must depend on the shape of the initial spectrum of pre-existing electrons
\citep[e.g.,][]{kangryu15}.
For the preshock electron population of a power-law form, $f_{\rm up}\propto p^{-s}$,
for example,
the ensuing radio spectrum at the shock position can have 
 $\alpha_{\rm sh}\le(s-3)/2$ or $\alpha_{\rm sh}\le (q-3)/2$,
depending on $s$, $q$, and observation frequencies \citep{kangryu16}.
Hence, in the re-acceleration model,
the Mach number of the putative shock cannot be inferred directly from the radio spectral index,
based on the expectation of the simple version of DSA model.
This may explain the fact that in some radio relics such as the Toothbrush relic, the radio Mach number, $M_{\rm radio}$, estimated from the radio spectral index
does not agree with the X-ray Mach number, $M_{\rm X-ray}$, estimated from the discontinuities in X-ray observations \citep[e.g.,][]{akamatsu13}

\subsection{Width of Radio Relics}

The cluster RX J0603.3+4214 that hosts the Toothbrush relic is located at the redshift $z=0.225$ \citep{vanweeren12}.
According to the recent observational study of \citet{vanweeren16}, 
FWHMs of B1 relic at 150~MHz and 610~MHz are about 140~kpc and 110~kpc, respectively,
and the spectral index between these two frequencies,
$\alpha_{0.15}^{0.61}$ increases from 0.8 at the northern edge of B1
to 1.9 at $\sim200$~kpc toward the cluster center.
The width of B1 relic at 610~MHz is
about 2 times larger than the FWHM $\sim 55$~kpc of the Sausage relic at 630~MHz \citep{vanweeren10}.

In the postshock region, accelerated electrons loose energy via synchrotron emission and
inverse Compton scattering off the cosmic background radiation in the following time scale:
\begin{equation}
t_{\rm rad} (\gamma_e) =  9.8\times 10^{7} \yr
\left[ {(5\muG)^2} \over {B^2+B_{\rm rad}^2}  \right]^{2}
 \hskip-3pt\left({\gamma_e \over 10^4 }\right)^{-1}
\label{trad}
\end{equation}
\citep{kang11}.
For low energy electrons,
the physical width of the postshock volume of radio-emitting electrons 
is simply determined by the advection length, $\Delta l_{\rm adv}\approx u_2 \cdot t_{\rm age}$,
where $u_2$ is the downstream flow speed. 
For high energy electrons, on the other hand, it becomes the cooling length, 
$\Delta l_{\rm cool}(\gamma_e)\approx u_2 \cdot t_{\rm rad}(\gamma_e)$.
So the width of the radio-emitting shell of electrons with $\gamma_e$
is $\Delta l(\gamma_e)={\rm min}[\Delta l_{\rm cool}(\gamma_e),\Delta l_{\rm adv}]$. 

In order to connect the spatial distribution of electrons with that of radio emission,  
one can use the fact that 
the synchrotron emission from mono-energetic electrons with $\gamma_{\rm e}$ peaks around
\begin{equation}
\nu_{\rm peak} 
\approx 0.63\ {\rm GHz} \cdot \left({B\over {5\muG} }\right) \left({\gamma_e \over 10^4}\right)^2,
\label{nupeak}
\end{equation}
along with the relation between the observation frequency and the source frequency,
$\nu_{\rm obs}= \nu_{\rm peak}/(1+z)$.
For low frequency radio emission emitted by uncooled, low energy electrons,
the width of radio relics becomes similar to the advection length: 
\begin{equation}
\Delta l_{\rm low}\approx 
100~{\rm kpc}\cdot u_{2,3}\cdot \left({{t_{\rm age}}\over {100~{\rm Myr}}}\right),
\end{equation}
where $u_{2,3}=u_2/10^3 \kms$. 

For high frequency emission radiated by cooled high energy electrons, 
the relic width would be similar to the cooling length \citep{kangryu15}:
\begin{eqnarray}
\Delta l_{\rm high}
\approx 100\ {\rm kpc} \cdot W\cdot u_{2,3}\cdot Q\cdot
\left[{{\nu_{\rm obs}(1+z) \over {0.63{\rm GHz}}} }\right]^{-1/2}.
\label{lwidth}
\end{eqnarray}
Here the factor $W\sim 1.2-1.3$ takes account 
for the fact that the spatial distribution of synchrotron emission at $\nu_{\rm peak}$ is somewhat broader than
that of electrons with the corresponding $\gamma_e$, because more abundant, lower energy electrons
also make contributions \citep{kang15a}.
The factor $Q$ is defined as
\begin{equation}
Q(B,z)\equiv \left[ { {(5\muG)^2} \over {B^2+B_{\rm rad}(z)^2}}\right] \left({B \over 5 \muG}\right)^{1/2},
\label{qfactor}
\end{equation}
where again $B$ and $B_{\rm rad}$ must be expressed in units of microgauss.
Figure 1 shows that $Q$ evaluated for $z=0.225$ peaks at $B\approx 2.8\muG$.
Note that for a given value $Q$ there are two possible values of $B$ that satisfies Equation (\ref{qfactor}).

\begin{figure}[t]
\centering
\includegraphics[width=70mm]{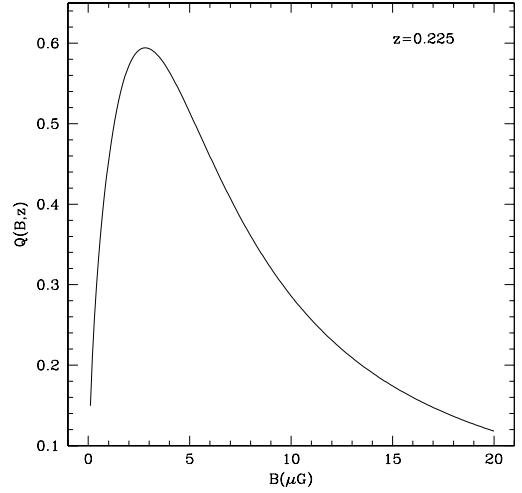}
\caption{The factor $Q(B,z)$ for $z=0.225$ given in Equation (\ref{qfactor}).
}
\end{figure}

\begin{figure*}[t]
\centering
\includegraphics[width=140mm]{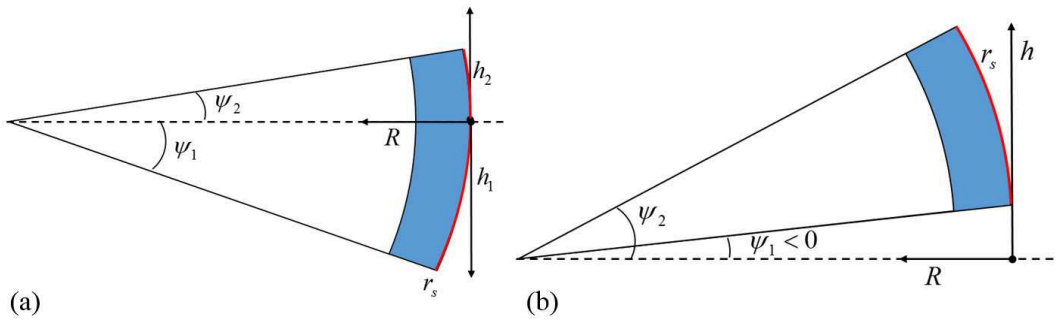}
\caption{Two geometrical configurations of the radio-emitting volume (marked in blue) 
downstream of a spherical shock at $r_s$ (marked in red) for the Toothbrush relic.
}
\end{figure*}

For B1 relic in the Toothbrush cluster, 
the estimated downstream speed is $u_2\approx 1.1\times 10^3\kms$,
if we take the following observation data: $kT_1\sim 6-7$ keV and $M_s\sim 1.3-1.5$ \citep{vanweeren16}. 
So if we adopt $W\approx1.3$, $u_{2,3}\approx 1.1$, and $t_{\rm age}\approx 110$~My,
then the width of low-frequency radio emission in a {\it planar} shock becomes 
$\Delta l_{\rm low}\approx 120$~kpc.
On the other hand, at high frequencies, for example, at $\nu_{\rm obs}=630$~MHz
it becomes only $\Delta l_{\rm high}\approx 78$~kpc for the largest value $Q_{\rm max}\approx0.6$
(with $B\approx 2.8\muG$). 

As shown in Figure 2, the projected relic width of a spherical shell can be larger 
than $\Delta l_{\rm low}$ or $\Delta l_{\rm high}$,
if the arc-like partial shell (marked in blue) is tilted with respect to the sky plane (dashed line). 
On the other hand, if the shell were to be viewed almost face-on,
the spectral index $\alpha$ would be roughly constant across the width of the relic, 
which is in contradiction with observations 
\citep{vanweeren16}.
Later we will consider models with the various projection angles, $-5^{\circ}\le \psi_1\le +15^{\circ}$
and $0^{\circ}\le\psi_2\le +35^{\circ}$.

\section{Numerical Calculations}

The numerical setup and physical models for DSA simulations were described in details in \citet{kang15b}.
So only basic features are given here.

\subsection{DSA Simulations for 1D Spherical Shocks}

We follow time-dependent diffusion-convection equation
for the pitch-angle-averaged phase space distribution function
for CR electrons, $f_e(r,p,t)=g_e(r,p,t)p^{-4}$, in the one-dimensional (1D) 
spherically symmetric geometry:

\begin{eqnarray}
{\partial g_e\over \partial t}  + u {\partial g_e \over \partial r}
= {1\over{3r^2}} {{\partial (r^2 u) }\over \partial r} \left( {\partial g_e\over
\partial y} -4g_e \right)  \nonumber\\
+ {1 \over r^2}{\partial \over \partial r} \left[ r^2 D(r,p)
{\partial g_e \over \partial r} \right]
+ p {\partial \over {\partial y}} \left( {b\over p^2} g_e \right),
\label{diffcon}
\end{eqnarray}
where $u(r,t)$ is the flow velocity, $y=\ln(p/m_e c)$, $m_e$ is the electron mass, and $c$ is
the speed of light \citep{skill75}.
Here $r$ is the radial distance from the center of the spherical coordinate,
which assumed to coincide with the cluster center.
We assume a Bohm-like spatial diffusion coefficient, $D(r,p)\propto p/B$. 
The cooling term $b(p)=-dp/dt= - p/t_{\rm rad} $ accounts for electron synchrotron and iC losses.
The test-particle version of CRASH (Cosmic-Ray Amr SHock) code in a comoving 1D spherical grid is used to
solve Equation (\ref{diffcon}) \citep{kj06}.

\subsection{Models for Magnetic Fields}

We consider simple yet physically motivated models for the postshock magnetic fields as in \citet{kang15b}:
(1) the magnetic field strength across the shock transition is assumed to increase 
due to compression of the two perpendicular components,
\begin{equation}
B_2(t)=B_1 \sqrt{1/3+2\sigma(t)^2/3}, 
\label{b2}
\end{equation}
where $B_1$ and $B_2$ are the preshock and postshock magnetic field strengths,
respectively, and $\sigma(t)=\rho_2/\rho_1$ is the density compression ratio across the shock.
(2) for the downstream region ($r<r_s$), the magnetic field strength is assumed to scale with the 
gas pressure:
\begin{equation}
B_{\rm dn}(r,t)= B_2(t) \cdot [P_g(r,t)/P_{g,2}(t)]^{1/2},
\label{bd}
\end{equation}
where $P_{g,2}(t)$ is the gas pressure immediately behind the shock.

\begin{figure*}[t!]
\vskip -0.5cm
\centering
\includegraphics[trim=2mm 2mm 2mm 2mm, clip, width=150mm]{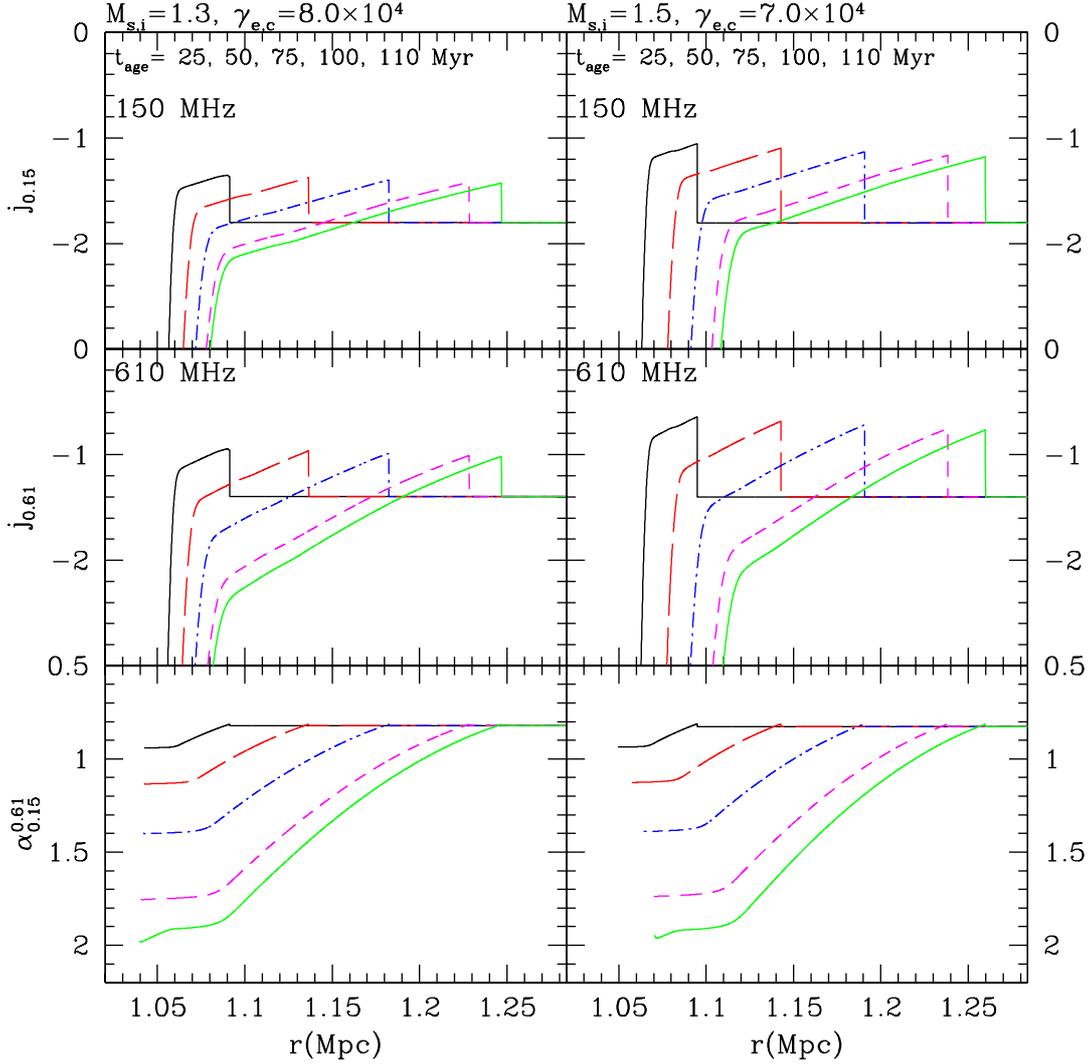}
\caption{Time Evolution of the synchrotron emissivity, $j_{\nu}(r)$ at 150~MHz (top panels) and 610~MHz (middle), 
and the spectral index,
$\alpha_{0.15}^{0.61}$ between 150 and 610~MHz (bottom) plotted as a function of the radial distance from the cluster center, $r(\rm Mpc)$,
in the two models with the initial Mach number $M_{s,i}=1.3$ (left panels) and $M_{s,i}=1.5$ (right panels).
The results at 25 (black solid lines), 50 (red long dashed), 75 (blue dot-dashed), 100 (magenta dashed), and 110~Myr (green solid) are presented.
}
\end{figure*}

\begin{figure*}[t!]
\vskip -3.3cm
\centering
\includegraphics[trim=2mm 2mm 2mm 2mm, clip, width=150mm]{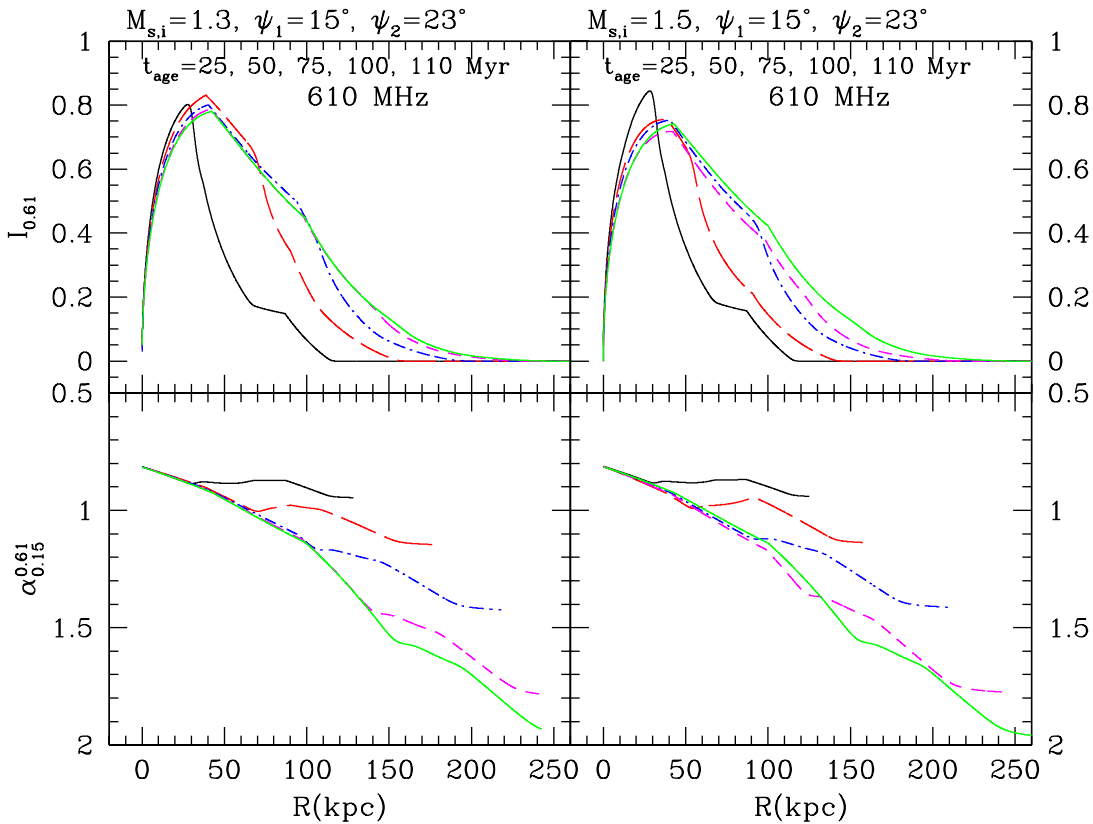}

\caption{Time Evolution of the surface brightness $I_{\nu}$ at 610~MHz (top panels) and the
spectral index, $\alpha_{0.15}^{0.61}$ between 150 and 610~MHz plotted as a function of the projected distance behind the shock, $R(\rm kpc)$,
in the two models with the initial Mach number $M_{s,i}=1.3$ (left panels) and $M_{s,i}=1.5$ (right panels).
The projection angles are assume to be $\psi_1=15^{\circ}$ and $\psi_2=23^{\circ}$.
The results at 25 (black solid lines), 50 (red long dashed), 75 (blue dot-dashed), 100 (magenta dashed), and 110~Myr (green solid) are presented.
}
\end{figure*}

\begin{figure*}[t!]
\vskip -0.5cm
\centering
\includegraphics[trim=2mm 2mm 2mm 2mm, clip, width=150mm]{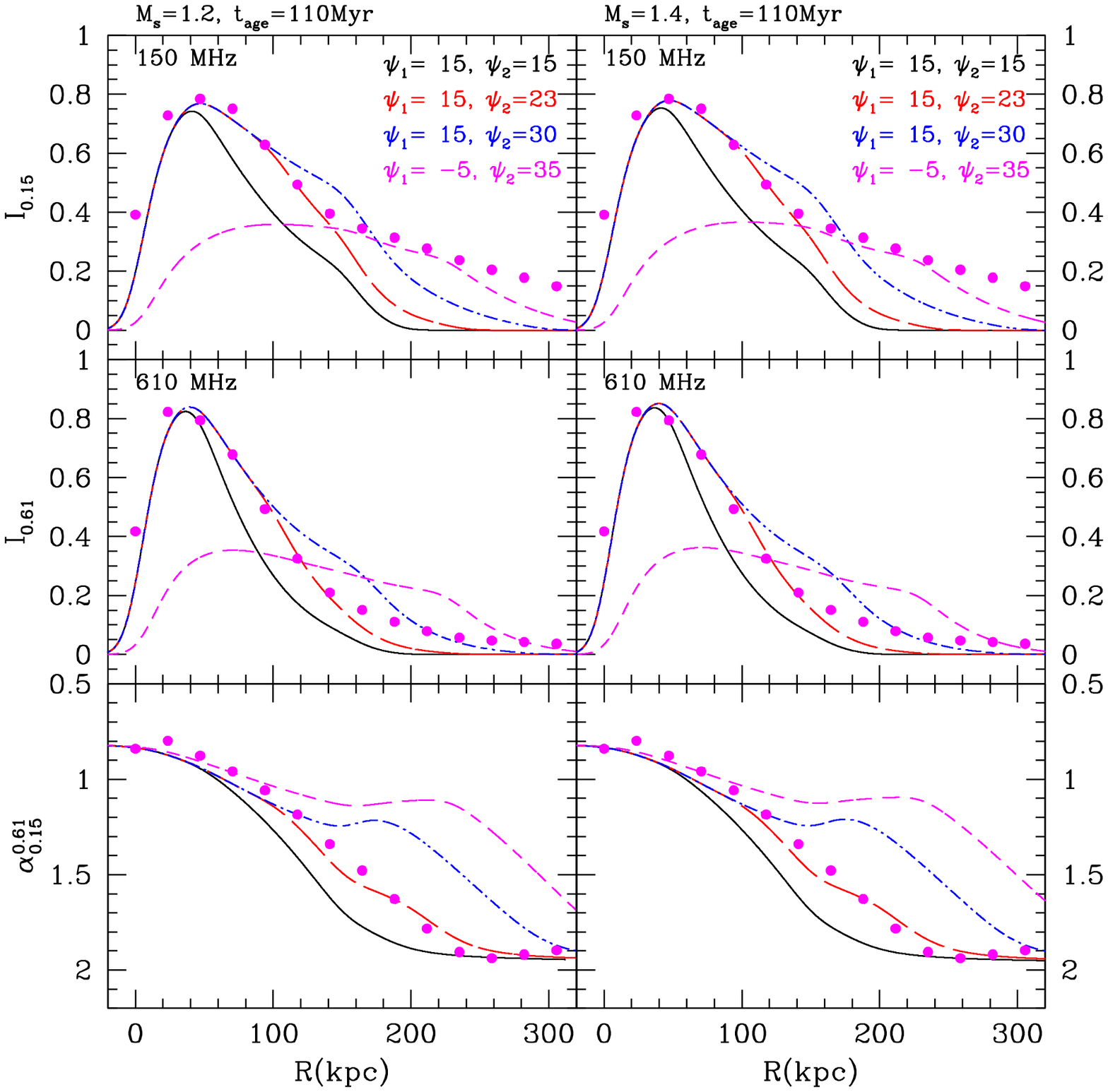}
\caption{Surface brightness $I_{\nu}$ at 150~MHz (top panels) and at 610~MHz (middle panels),
and the
spectral index $\alpha_{0.15}^{0.61}$ between the two frequencies (bottom panels)
plotted as a function of projected distance behind the shock, $R(\rm kpc)$,
for the shock models with $M_s=1.2$ (left panels) and $M_s=1.4$ (right panels) at $t_{\rm age}=110$~Myr .
The projection angles, $\psi_1$ and $\psi_2$, for different curves are given in the plot.
Gaussian smoothing with $6.5"$ resolution ($= 23.5$~kpc) is applied in order to emulate radio observations.
The magenta dots are the observational data of \citet{vanweeren16}.
}
\end{figure*}

\subsection{Preshock Electron Population}

Several possible origins for pre-existing, relativistic electron populations in the ICMs can be considered:
(1) old remnants of radio jets from AGNs (radio ghosts), (2) electron populations that were accelerated by 
previous shocks, and
(3) electron populations that were accelerated by turbulence during merger activities 
\citep[e.g.,][]{kangryu15}.
In the case of the Toothbrush relic, \citet{vanweeren16} suggested that the AGN located at the southwestern end 
of B1 relic might be the candidate source that supplies relativistic plasmas to this relic.

Here we assume that a preshock population of electrons has a power-law spectrum 
with exponential cutoff as follows: 
\begin{equation}
f_{\rm up}(p) \propto p^{-s} \exp \left[ - \left({\gamma_e \over \gamma_{e,c}} \right)^2 \right],
\label{fexp}
\end{equation}
where $s=4.6$ is chosen to match the observed shock index, $\alpha_{\rm sh}\approx 0.8$. 
We consider several models with a range of the cutoff Lorentz factor, $\gamma_{e,c}= 2-10\times 10^4$,
and find that $\gamma_{e,c}\sim 7-8\times 10^4$ produces the best match to the observations of
both \citet{stroe16} and \citet{vanweeren16}.

Note that this preshock electron population could be detected in radio, as can be seen in Figure 3, 
since $\gamma_{e,c}$ is quite high.
However, no radio fluxes have been detected north of the northern edge of B1 (the shock location) \citep{vanweeren16}.
Since the candidate AGN is located about 200~kpc downstream of the shock location,
it is possible that the shock has almost sweeps through the upstream region with these
preshock electrons emitted by the AGN.

In order to isolate the effects of the pre-existing population,
the {\it in situ} injection at the shock is turned off in the DSA simulations presented here.

\subsection{Shock parameters}

The shock parameters are chosen to emulate roughly 
the shock associated with B1 relic in the Toothbrush cluster. 
According to \citet{vanweeren16},
$kT_1=8.3_{-2.4}^{+3.2}$ keV and $kT_2=8.2_{-0.9}^{+0.7}$ keV across B1,
so the postshock temperature is better constrained compared to the preshock temperature.
Considering that $\Delta l_{\rm high}$ is smaller the observed width ($\sim 110$~kpc) at 610~MHz,
we choose a higher value of $kT_2=8.9$ keV in order to maximize $\Delta l_{\rm high}$.
Then we consider two values of the initial Mach number,
which determine the preshock temperature and the initial shock speed as follows:

(1) $M_{s,i}=1.3$, $kT_1=6.9$ keV, $u_{s,i}=1.8\times10^3\kms$, and $\gamma_{e,c}=8.0\times10^4$.

(2) $M_{s,i}=1.5$, $kT_1=5.9$ keV, $u_{s,i}=1.9\times10^3\kms$,  and $\gamma_{e,c}=7.0\times10^4$.

The spherical shock slows down to $M_s\approx 1.2$ in the $M_{s,i}=1.3$ model
and $M_s\approx 1.4$ in the $M_{s,i}=1.5$ model at the shock age of $t_{\rm age}\approx 110$~Myr.
The preshock magnetic field strengths is assumed to be $B_1= 2\muG$, resulting in the
postshock strength, $B_2=2.5-2.8\muG$.

The density of the background gas in the cluster outskirts is assumed to decrease as 
$\rho_{\rm up}=\rho_0(r/r_{s,i})^{-2}$.
This corresponds to the so-called beta model for isothamal ICMs,
$\rho(r)\propto [ 1+ (r/r_c)^2 ]^{-3\beta/2}$ with $\beta \sim 2/3$ \citep{sarazin88}.

\section{RESULTS OF DSA SIMULATIONS}

Figure 3 shows the synchrotron emissivity, $j_{\nu}(r)$ at 150~MHz and 610~MHz, and the spectral index,
$\alpha_{0.15}^{0.61}$ between the two frequencies
in the models with $M_{s,i}=1.3$ and $M_{s,i}=1.5$.
It illustrates how the downstream radio-emitting region broadens in time as the spherical shock moves outward
away from the cluster center, and how the radio spectrum steepens behind the shock further in time.
In the model with $M_{s,i}=1.3$, the emissivities at both frequencies increase by a factor of $2.4-2.7$ across the shock,
while they increase by a factor of $4.3-5.6$ in the model with $M_{s,i}=1.5$.
Again note that no radio fluxes have been detected in the upstream region of B1 relic \citep{vanweeren16}.
In that regard, the $M_{s,i}=1.5$ model might be preferred, since it has a larger amplification factor
across the shock.

The emissivity $j_{0.61}(r)$ decreases much faster than $j_{0.15}(r)$ behind the shock,
since the cooling length of higher energy electrons is shorter than that of lower energy electrons.
The spectral index at the shock is $\alpha_{0.15}^{0.61}\approx 0.8$, since the
slope of the pre-existing electron population is $s=4.6$ in Equation (\ref{fexp}).
It increases to $\alpha_{0.15}^{0.61}\approx 1.9$ at the far downstream regions, 
which is in good agreement with observations.

Note that in the shock rest frame the postshock flow speed, $u_{\rm dn} (r,t)$, increases behind spherically expanding shocks, 
so the advection length is larger than that estimated for planar shocks.
According to Figure 3, at the shock age of 110~My, for instance, 
the advection lengths become $\sim150-160$~kpc,
while $\Delta l_{\rm low}\approx 120$~kpc estimated for planar shocks.

\subsection{Surface Brightness and Spectral Index Profiles}

The radio surface brightness, 
$I_{\nu}(R)$, is calculated from the emissivity $j_{\nu}(r)$ by adopting the same geometric volume of radio-emitting electrons
as in Figure 1 of \citet{kang15b}.
Here $R$ is the distance behind the projected shock edge in the plane of the sky,
while $r$ is the radial distance from the cluster center.
As shown in Figure 2, the various projection angles $\psi_1+\psi_2=30^{\circ}-45^{\circ}$ are considered
in order to explore the projection effects.
For example, if both $\psi_1> 0$ and $\psi_2> 0$, the surface brightness is calculated as follow:
\begin{equation}
I_{\nu}(R)= \int_0^{h_{\rm 1, max}} j_{\nu}(r) d {\it h_1}+ \int_0^{h_{\rm 2, max}} j_{\nu}(r) d {\it h_2}, 
\label{SB}
\end{equation}
where $h$ is the path length along line of sights. 
Note that the radio flux density, $S_{\nu}$, can be obtained by convolving $I_{\nu}$ with a telescope beam as
$S_{\nu}(R) \approx I_{\nu}(R) \pi \theta_1 \theta_2 (1+z)^{-3}$,
if the brightness distribution is broad compared to the beam size of $\theta_1 \theta_2$.
In addition, the spectral index, $\alpha_{0.15}^{0.61}$ is calculated from the projected
$I_{\nu}(R)$ at 150 and 610~MHz.

Figure 4 shows how $I_{\nu}(R)$ at 610~MHz and $\alpha_{0.15}^{0.61}$ evolve during 110~Myr
in the two shock models.
The projection angles, $\psi_1=15^{\circ}$ and $\psi_2=23^{\circ}$,
are chosen to match the observed profiles of $I_{\nu}(R)$ and $\alpha_{0.15}^{0.61}$,
as shown in Figure 5.
The shock has weakened to $M_s\approx 1.2$ in the $M_{s,i}=1.3$ model
and $M_s\approx 1.4$ in the $M_{s,i}=1.5$ model at 110~Myr.
In the early stage the relic width increases roughly with time.
But it asymptotes to the cooling lengths in time, since the cooling of
high energy electrons begins to control the width at later time.

The gradient of $\alpha_{0.15}^{0.61}$ increases in time as postshock electrons loose energies.
At $t_{\rm age}=110$~My, the spectral index increases from $\alpha_{0.15}^{0.61}\approx 0.8$ at the shock location
to $\alpha_{0.15}^{0.61}\approx 1.9$ about 230~kpc downstream of the shock in the two models.
Note that the evolution of $\alpha_{0.15}^{0.61}$ depends on the cutoff energy of the
pre-existing electron populations.
In order to obtain the observed profile that steepens from 0.8 to 1.9 over 230~kpc,
we should choose $\gamma_{e,c}\approx 8.0\times 10^4$ for $M_{s,i}=1.3$ model
and $\gamma_{e,c}\approx 7.0\times 10^4$ for $M_{s,i}=1.5$ model.

We find that the best fits to the observations of \citet{vanweeren16} are obtained
at 110~Myr for both models.
Figure 5 shows $I_{\nu}(R)$ at 150 and 610~MHz and $\alpha_{0.15}^{0.61}$ at $t_{\rm age}=110$~Myr
for different sets of $\psi_1$ and $\psi_2$.
Note that Gaussian smoothing with $6.5"$ resolution ($= 23.5$~kpc) is applied 
in order to emulate radio observations.
The scales of $I_{\nu}(R)$ are rescaled in arbitrary units.
If we were to attempt to match the observed profile of $I_{0.15}$ up to $\sim 300$~kpc, 
then one of the projection angles should be larger than $35^{\circ}$,
as can be seen in the top panels.
With such projection angles, however, the predicted profile of $\alpha_{0.15}^{0.61}$
would become much flatter than the observed profile. 
Since the observed radio fluxes could be contaminated by the
halo components, we try to fit the observed profiles only for $R<150$~kpc.
As can be seen in the figure,
the model with $\psi_1=15^{\circ}$ and $\psi_2=23^{\circ}$ gives the best match to the observations.
At 110~My the shock radius is $r_s(t)\approx 1.25-1.26$~Mpc (see Figure 3).
Note that in fitting the profile of $I_{\nu}(R)$ with $\psi_1=15^{\circ}$ and $\psi_2=23^{\circ}$,
 $r_s(t)(1-\cos\psi_1)\approx 43$~kpc is related with the peak position at $\sim 47$~kpc,
while $r_s(1-\cos\psi_2)\approx 100$~kpc is related with the relic width of $\sim 110-140$~kpc.

\subsection{Integrated Spectrum}

As described in Section 2.1, 
the volume integrated radiation spectrum of a typical cluster shock with $t_{\rm age}\sim 100$~Myr
is expected to steepen from $\alpha_{\rm sh}$ to $\alpha_{\rm sh}+0.5$ gradually over $0.1 - 10$ GHz.
In the re-acceleration model, however, it also depends on the spectral shape of the preshock electron 
population. 
So the spectral curvature of the observed spectrum can be reproduced by adjusting the set of
model parameters, i.e., $M_s$, $B_1$, $t_{\rm age}$, $s$ and $\gamma_{e,c}$ in our models.

Figure 6 shows the time evolution of the integrated spectrum, $J_{\nu}$, during 110~Myr
in the two shock models.
The magenta dots show the observational data taken from Table 3 of \citet{stroe16},
which are rescaled to fit the simulated spectrum near 1~GHz by eye. 
We find that $\gamma_{e,c}=7-8\times 10^4$ is needed in order to reproduce $J_{\nu}$ both
near 1~GHz and $16-30$~GHz simultaneously.
At 110~Myr (green solid lines) the predicted fluxes are slightly higher than the observed fluxes  
for $0.1-1$~GHz, while the predicted fluxes are in good agreement with
the observed fluxes for $1-30$~GHz.
We can conclude that the model spectra agree reasonably well with the observed spectrum
of B1 relic reported by \citet{stroe16}.

\begin{figure*}[t!]
\vskip -1.5cm
\centering
\includegraphics[trim=2mm 3mm 2mm 2mm, clip, width=150mm]{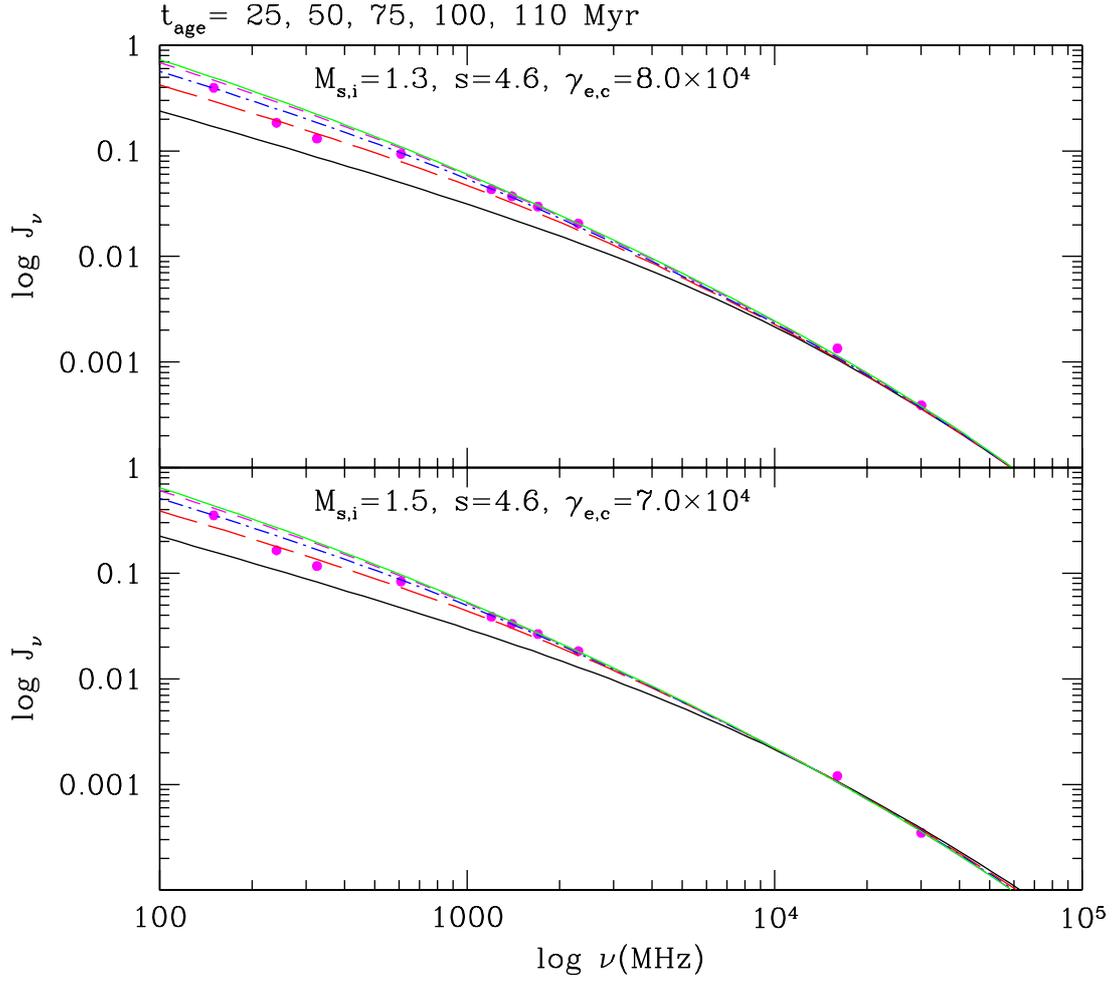}
\caption{Time evolution of volume-integrated radio spectrum for the shock models
with $M_{s,i}=1.3$ (top panels) and $M_{s,i}=1.5$ (bottom) at $t_{\rm age}=$ 25 (black solid lines), 50 (red long-dashed), 
75 (blue dot-dashed), 100 (magenta dashed), and 110~Myr (green solid).
The magenta dots are the observational data taken from Table 3 of \citet{stroe16}.
}
\end{figure*}

\section{Summary}

We have performed time-dependent, DSA simulations of one-dimensional, spherical
shocks, which sweep through the ICM thermal plasma with
a population of relativistic electrons ejected from an AGN.
In order to reproduce the observational data for B1 (head portion) of the Toothbrush relic,
we adopt the following parameters: $kT_1\approx 5.9-6.9$~keV, $M_{s,i}=1.3-1.5$, 
$u_{s,i}=1.8-1.9\times 10^3 \kms$, and $B_1=2\muG$
\citep{vanweeren16}.

In addition, we assume that the preshock gas contains a pre-existing electron population 
of a power-law spectrum with the slope $s=0.6$ and exponential cutoff at $\gamma_{e,c}=7-8\times 10^4$
as given in Equation (\ref{fexp}).
The power-law slope $s$ is chosen to match the observed radio spectral index, $\alpha_{\rm sh}\approx 0.8$,
while $\gamma_{e,c}$ is adopted to match the integrated radio spectrum of B1 relic for $1-30$~GHz
\citep{stroe16}.
Note that the preshock region could be radio-luminous with the pre-existing electrons with such high $\gamma_e$.
In the DSA models considered here,
the synchrotron emissivity increases across the shock by a factor of $2.4-2.7$ in the $M_{s,i}=1.3$ model
and by a factor of $4.3-5.6$ in the $M_{s,i}=1.5$ model.

After 110~Myr, the spherical shock slows down to $M_s\approx 1.2-1.4$ with the postshock magnetic field strength
$B_2\approx 2.5-2.8 \muG$.
At this stage, the spatial profiles of
the surface brightness, $I_{\nu}$, and the radio spectral index, $\alpha_{0.15}^{0.61}$,
and the integrated radio spectrum, $J_{\nu}$, become consistent with the observations of B1 relic
 \citep{vanweeren16}.
Considering that the candidate source AGN is located about 200~kpc downstream of the shock (the northern edge of B1
relic) and there is no detectable radio fluxes in the preshock region,
we suspect that the shock might have swept through the region permeated by relativistic plasma ejected from
the source AGN.

The simple estimations for the widths of the downstream volume of radio emission behind a planar shock
are about $\Delta l_{\rm low}\approx 120$~kpc at 150~MHz and $\Delta l_{\rm high}\approx 78$~kpc at 610~MHz
at the shock age of 110~Myr.
On the other hand, the width of the projected surface brightness of a spherical shell
depends on the projection angles as shown in Figure 2.
The observed profiles of radio fluxes at both frequencies can be modeled 
with the projection angles, $\psi_1=15^{\circ}$ and $\psi_2=23^{\circ}$,
as shown in Figure 5. 
So the arc-like radio-emitting shell is likely to be tilted slightly with respect to the sky plane as shown
in the case (a) of Figure 2.

\acknowledgments{
This research was supported by Basic Science Research Program through 
the National Research Foundation of Korea (NRF) funded by the Ministry 
of Education (2014R1A1A2057940).
The author thanks R.~J. van Weeren for providing the radio flux data for the Toothbrush relic
published in \citet{vanweeren16}.
}


\end{document}